\documentclass[prl,twocolumn,superscriptaddress]{revtex4-2} 
\usepackage{mathrsfs}
\usepackage{amsfonts}
\usepackage{amsmath}
\usepackage{amssymb}
\usepackage{graphicx,subfigure}
\usepackage{bm}
\usepackage{color}
\usepackage[normalem]{ulem}
\usepackage{dcolumn} 
\usepackage[table]{xcolor}
\usepackage{float}
\usepackage{tabularx}
\usepackage{hyperref}
\newcommand{\ket}[1]{|#1\rangle}
\newcommand{\bra}[1]{\langle #1|}

\begin{document}

\title{Comment on ``Evolution Operator Can Always Be Separated into the Product of Holonomy and Dynamic Operators''} 

\author{Adam Fredriksson}
\affiliation{Department of Physics and Astronomy, Uppsala University, Box 516, SE-751 20 Uppsala, Sweden}

\author{Erik Sj\"oqvist}
\affiliation{Department of Physics and Astronomy, Uppsala University, 
Box 516, SE-751 20 Uppsala, Sweden}
\email{erik.sjoqvist@physics.uu.se}

\date{\today}

\maketitle

In a recent Letter \cite{yu23}, Yu and Tong derived an expression for a generic time 
evolution operator acting on a subspace, claimed to constitute a general separation 
into a product of a holonomy operator and a dynamic operator. This result appears 
to be in conflict with prior work by Anandan \cite{anandan88}, in which the time 
evolution was found not to separate into a holonomic and a dynamic part in general. 
In this Comment, we show that the claim in Ref.~\cite{yu23} that the time evolution 
operator always can be written as a product of a holonomy operator and a dynamic 
operator is false, as it is based on a circular use of the time evolution operator.

The key finding of Ref.~\cite{yu23} is that given a Hilbert space $\mathscr{H}$ of some 
quantum system and the $\ell$ dimensional ($\ell \leq \dim \mathscr{H}$) subspace 
spanned by $\{\ket{\psi_{j}(t)}\}_{j = 1}^{\ell}$, where each $\ket{\psi_{j}(t)}$ is a solution 
of the Schr\"odinger equation with Hamiltonian $H(t)$, the time evolution operator 
$U(t,0) = \ket{\psi_{j}(t)}\bra{\psi_{j}(0)}$ satisfies the differential equation 
\begin{eqnarray}
\dot{U}(t,0) = \dot{P}(t) U(t,0) + U(t,0) F(t,0),
\label{eq:yte}
\end{eqnarray}
where $P(t)\equiv\ket{\psi_{j}(t)}\bra{\psi_{j}(t)}$ is the projector on the subspace, 
$F(t,0) \equiv F_{jk}(t)\ket{\psi_{j}(0)}\bra{\psi_{k} (0)}$ with 
$F_{jk}(t)\equiv -i\bra{\psi_{j}(t)} H(t)\ket{\psi_{k} (t)}$, we use Einstein's summation 
convention, i.e., repeated indices are implicitly summed, and we put $\hbar = 1$.
Formally, Eq.~\eqref{eq:yte} implies that
\begin{eqnarray}
U(t,0) & = & 
{\mathcal{P}}e^{\int_0^t \dot{P} (\tau) d\tau} 
P(0) \bar{\mathcal{T}} e ^{\int_{0}^{t}  F(\tau,0) d\tau} , 
\label{eq:yte_solution}
\end{eqnarray} 
which is the expression for $U(t,0)$ found in Ref.~\cite{yu23} that was claimed to 
imply a separation into a product of a holonomy operator and a dynamic operator, 
given by a path ordered exponential  ${\mathcal{P}}e^{\int_0^{t} \dot{P} (\tau) d\tau} P(0)$ 
and a reverse time ordered exponential $P(0)\bar{\mathcal{T}} e ^{\int_{0}^{t}  F(\tau,0) d\tau}$, 
respectively. 

To understand why Eq.~\eqref{eq:yte_solution} does not imply the claimed separation, 
one should note that $F(t,0)$ in fact depends on the time evolution operator itself: 
\begin{eqnarray}
F(t,0) & = & -i\ket{\psi_j (0)} \bra{\psi_{j}(t)} H(t)\ket{\psi_{k} (t)} \bra{\psi_{k} (0)} 
\nonumber \\ 
 & = & -iU^{\dagger}(t,0)H(t)U(t,0) .    
 \label{eq:F}
\end{eqnarray}
This observation, which seems to be overlooked in Ref.~\cite{yu23}, allows us to 
write Eq.~\eqref{eq:yte_solution} as  
\begin{eqnarray}
U(t,0) = 
\mathcal{P}e^{\int_0^{t} \dot{P} (\tau) d\tau} P (0) 
\bar{\mathcal{T}}
e ^{-i\int_{0}^{t} U^{\dagger} (\tau,0) H(\tau) U(\tau,0) d\tau} . 
\nonumber \\
\label{eq:ytesol}  
\end{eqnarray}  
Evidently, the `dynamic' operator, the last factor on the right-hand side, contains 
contributions from the holonomy operator via $U(t,0)$ in a circular manner. In other 
words, if one insists on calling Eq.~\eqref{eq:yte_solution} a separation into a product 
of a holonomy operator and a dynamic operator, it follows, by Eq.~\eqref{eq:ytesol}, 
that the `dynamic' operator depends non-trivially on the holonomy operator. Thus, 
contrary to the claim in Ref.~\cite{yu23}, Eq.~\eqref{eq:yte_solution} does not 
constitute a general separation of $U(t,0)$ into a holonomy operator and dynamic 
operator.

\end{document}